\documentclass{article}%
\input epsf
\usepackage{amsmath}%
\usepackage{amsfonts}%
\usepackage{amssymb}%
\usepackage{graphicx}
\newcommand{\be}[1]{
\begin{eqnarray}\label{#1}}
\newcommand{\ee}{\end{eqnarray}}
\newcommand{\ci}[1]{\cite{#1}}
\newcommand{\re}[1]{(\ref{#1})}

\newcommand{\tr}{\mbox{\rm tr}}
\newcommand{\ba}{\begin{array}}
\newcommand{\ea}{\end{array}}

\newcommand{\partialboth}{\stackrel{\leftrightarrow}{\partial}}

\begin{document}
%\rightline{RUB-TPII-02/02}

\begin{center}
{\Large
%Breakdown    !changed
Difficulties
of chiral expansion for parton distributions}\\[0.5cm]

 N. Kivel$^{a,b}$, M.V.~Polyakov$^{a,b}$\\[0.3cm]

\footnotesize\it $^a$ Petersburg
Nuclear Physics Institute, Gatchina, St. Petersburg 188350,
Russia\\
 \footnotesize\it $^b$ Institute for
Theoretical Physics II, Ruhr University Bochum, Germany
\\
%\vspace*{1cm}
%{\bf DRAFT,  \today}

\end{center}

\begin{abstract}
In the framework of the chiral perturbation theory ($\chi$PT)
we computed two- and three-loop corrections to the pion parton distributions
which posses $\delta$-function singularities at $x_{Bj}\to 0$.
From this calculation
%explicitly demonstrates                                         !changed
one can conclude
that in the region
of small $x_{Bj}\sim m_\pi^2/(4\pi F_\pi)^2$ standard
$\chi$PT
%breaks down   and one                                            !changed
needs in the resummation
of all orders in order to get correct description of the PDFs
at small-$x$ region.
We
% give                                         !changed
demonstrate an example of such resummation.
\end{abstract}

\section{Introduction}

Recently chiral perturbation theory ($\chi$PT) has been applied to
 parton distribution functions (PDFs)
\cite{sav,kiv02,man,che} in order to determined
the dependence of these quantities on the mass of Goldstone bosons
($m_\pi$) and/or external small momenta ($\xi$ and $t$ in the case
of generalized parton distributions (GPDs)). In contrast to
standard $\chi$PT, in this case one deals with hadronic matrix
elements of  non-local light-cone operators ($n^2=0$):
\begin{equation}
 O(\lambda)  =
 \bar{q}\left( {\textstyle \frac{1}{2}}\lambda~n\right)
\gamma_{+}\{1,\,{\textstyle \frac{1}{2}}\tau^a \}
q\left(  -{\textstyle \frac{1}{2}}\lambda~n\right)  .
\label{Oper}
\end{equation}
Therefore apart from
chiral expansion parameters ($m_\pi$, small external momenta) one
has the additional scale $\lambda$ - the distance characterizing the
non-locality of the operator.
 Interesting theoretical
question is to study the interplay of this additional scale with
the chiral expansion parameters. In the present paper we investigate
this problem for the quark parton distributions in the pion.

For the case of the quark operators there exist two
independent distributions corresponding to the two
possible values of the isospin $I$:
 $I=0$ defines the singlet distribution $Q(x)$ and
 $I=1$ non-singlet one $q(x)$, which are
normalized as:
\be{norm}
\int_{-1}^1 dx~q(x)=1,~\int_{-1}^1 dx~x\ Q(x)=M_{2}^{Q}~.
\ee
Here $M_2^Q$ is the momentum fraction carried by quarks.
For convenience, the explicit definitions of the PDFs in terms of
the matrix elements of the light-cone operators are presented
in the Appendix.

The standard chiral expansion of these PDFs implies:
\be{Qchpt}
Q(x)  & =Q^{(0)}(x)+a_{\chi}~ Q^{(1)}(x)+a_\chi^{2}~Q^{(2)}%
(x)+...,\\
q(x)  & =q^{(0)}(x)+a_\chi~ q^{(1)}(x)+a_\chi^{2}~q^{(2)}%
(x)+...,
\label{qchpt}%
\ee
where $a_\chi=\left(m_\pi/4\pi F_\pi~\right)^{2}$
($F_\pi\approx 93$~MeV is the pion decay constant) is the chiral
expansion parameter.
 Important observation is that in the above formulae one assumes that
the momentum fraction $x$ has no chiral power counting, i.e.
\begin{equation}
x\sim\mathcal{O}(a_\chi^{0}).
\end{equation}

The one-loop non-analytic contributions have been computed in
many papers \cite{sav,kiv02,che,man}. They read:
\be{Q1loop}
Q^{(1)}(x)  & =&0\times\ln\left[  1/a_\chi\right]+{\cal O}(a_\chi)  ,\\
\label{q1loop} q^{(1)}(x)  & =&\left\{
~q^{(0)}(x)-\delta(x)\right\}  \ln\left[ 1/a_\chi \right]+{\cal
O}(a_\chi). \ee
We observe that isovector PDF $q(x)$ has specific
$\delta$-singular behavior at small momentum fraction. This can be
understood as following: for some small values of the momentum
fractions
 $x\sim a_\chi$ the next-to-leading singular term formally is of the same order as
the leading term $a_\chi \delta(x)\sim {\cal O}(a_\chi^0)$.
Moreover, higher orders of $\chi$PT may posses even
more singular structures, like derivatives of the
$\delta-$function: $\delta^{(n)}(x)$.
This may indicate that in this situation the usual chiral expansion
is not valid anymore.
The question is:
How to write correctly the chiral expansion for $x\sim a_\chi$ ( equivalently
for the large  light-cone distance $\lambda \sim 1/a_\chi$) ?
Is the chiral expansion for this case controlled by the finite order of the standard
$\chi$PT?

It is clear that in order to construct the improved  chiral expansion for the
region $x\sim a_\chi$ one has to perform resummation of all $\delta$-singular terms.
Then such reordering of the $\chi$PT expansion  may lead to generation of the
some finite size function instead of singular $\delta(x)$ and therefore
provides more correct chiral description of the small$-x$ behavior.

Let us formulate the above suggestion more accurately.
We shall assume, that performing the chiral limit at fixed value of the
momentum fraction $x$ one obtains the standard chiral
expansions \re{Qchpt},\re{qchpt} which can be rewritten as a sum of the
regular $f^{reg}(x)$ and $\delta-$singular parts
\be{QchptD}
Q(x)  & =Q^{reg}(x)+\delta^\prime (x)D_1
+\delta^{\prime\prime\prime} (x) D_3+...=
Q^{reg}(x)+\sum_{i\geq 1,\, {\rm odd}}D_i\, \delta^{(i)} (x)\, ,
\\
q(x)  & =q^{reg}(x)+\delta (x)D_0+\delta^{\prime\prime} (x) D_2+...
=q^{reg}(x)+\sum_{i\geq 0,\, {\rm even}}D_i\, \delta^{(i)} (x)\, ,
\label{qchptD}%
\ee
where the regular contributions can be represented in the framework of $\chi$PT
by series similar to \re{Qchpt}, \re{qchpt} and
where each term in the expansions is represented by
some smooth function of the momentum fraction $x$.
The coefficients $D_i$ in front of derivatives of the $\delta-$functions
are some constants which also can  be expanded with respect to small parameters
$a_{\chi}$:
\begin{align}
 D_i=D_i^{(0)}+a_{\chi} D_i^{(1)}+a^2_{\chi}D_i^{(2)}+...,
 \label{Dchpt}
\end{align}
We shall assume that PDFs are smooth functions in the chiral limit, i.e., for all
values $i=0,1,...$
\begin{align}
D_i^{(0)}=0.
\end{align}
Then the singular terms can occur only from the corrections
induced by the loop diagrams
of the chiral perturbation theory.
The power of the derivatives is dictated by the symmetry properties of the given PDF.
Recall that singlet
(non-singlet) pion PDFs  can be understood as antisymmetrical (symmetrical) function
with respect to exchange $x\leftrightarrow -x$ being extended to the whole region
$-1<x<1$.

In order to check the assumption \re{QchptD} or \re{qchptD}
 it would be interesting to see that
 the singular
contributions $\sim$ $\delta^{(n)}(x)$ can occur
at least in the first few orders of $\chi$PT expansion.
In next section we compute such singular
contributions to the PDFs and find
the explicit  contributions  of the
type $ [a_\chi \ln(1/a_\chi)]^{n+1}\ \delta^{(n)}(x)$
for the cases $n=1,2$.
We shall see that the
mechanism of the generation of $\delta-$functions is quite general and
should work also at higher orders.
In the section Discussion we briefly discuss the main consequences which
follow from representations \re{QchptD} and \re{qchptD} and
possibilities to perform
the resummation of the singular contributions.
On the concrete example we demonstrate  that resummed PDF is given
by finite size function $f(x/a_\chi)$ which can be represented by the infinite sum
of the $\delta-$functions being expanded with respect to $a_\chi$.

\section{The singular $\delta-$function contributions in ChPT up to three loops}

Our goal is to investigate the higher orders of the chiral
expansion in order to establish the existence $\delta-$singular terms.
We shall accept following strategy: we restrict our consideration
only to leading non-analytic contributions. Such approach allows us to
perform all calculations in the framework of the leading order chiral Lagrangian
\be{WL}
 {\cal L}_{(2)}=\frac14 \, F^2_\pi\,  \tr\left(\,
 \partial^\mu U \partial_\mu U^\dag+\chi^\dag U+\chi U^\dag
 \right)\, ,
\ee
for the pion dynamics and leading order chiral representation for
the light-cone operators \re{Oper}. For convenience, all
necessary  technical details and definitions  are presented in the
Appendix. The accuracy of our approach, despite we are working
at higher orders, allows us to neglect the renormalization of the pion mass
and chiral couplings, hence below we always assume their physical values.

The calculation of the contributions of double chiral logarithms is
well-known subject and have been already discussed in the literature,
see for instance
\ci{col}. The main observation is that such terms occur due to the
$UV-$divergencies of the $\chi$PT diagrams and can be effectively computed
if one uses the basic property of  locality of the $UV-$counterterms.

\begin{figure}
[ptb]
\begin{center}
\includegraphics[
height=2cm
]%
{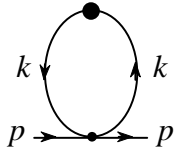}%
\end{center}
\caption{One-loop diagram which generates $\delta-$contribution to the pion PDF.
The large blob denotes two-pion non-local operator vertex generated by
 corresponding chiral Lagrangian for the light-cone operator}
\label{oneloop}
\end{figure}

We shall use dimensional regularization with space-time dimension $d=4-2\varepsilon$
in order to
compute the bare matrix elements. For the one-loop diagram the coefficient
in front of chiral logarithm is determined completely by the $UV-$pole, therefore
it is enough to compute only the $1/\varepsilon$-pole part of the graph.
This is clear from the general structure for
the unrenormalized matrix element:
\be{1loopme}
\int \frac{d\lambda}{2\pi}e^{-ip_+ x\lambda}
\left\langle p|O(\lambda )|  p\right \rangle |_{1\rm  loop}
\equiv
\langle O^{(1)}\rangle=
a_\chi\frac1\varepsilon \left(\frac{\mu^2_\chi}{m_\pi^2}\right)^{\varepsilon}
{D_0^{(1)}\,}\delta(x)
+ ...=
\\
a_\chi \left(\frac1\varepsilon +\ln\left[\frac{\mu^2_\chi}{m_\pi^2}\right]\right)
{D_0^{(1)}\,}\delta(x)+...\, .
\ee
where the chiral scale $\mu_\chi\sim 4\pi F_\pi$.
The interesting for us $\delta-$term is generated
from the diagram in Fig.\ref{oneloop}.
Let us describe shortly corresponding calculations.
The expression for the diagram is given
\be{onel}
G_1=i\varepsilon\lbrack abc]~a_{\chi}~
q^{(0)}(\beta)* \int dk~
\frac{ k_+ \delta(x p_+ -\beta k_+)}
{{m_\pi^2}\left[  k^{2}-m_\pi^{2}\right]^{2}}
\left[  -4(kp)+~...\right]  ,
\ee
where the dots in the brackets denote irrelevant (regular) contributions
from the four-pion vertex and by the asterisk we denote the convolution integral
with respect to $\beta$. We also neglected various terms with $\varepsilon$
which can provide only the finite terms and denote $dk\equiv d^{d}k/\pi^{d/2}$.
 Obviously, the integral in \re{onel} is quadratically divergent.
Moreover, from the structure of the denominator and numerator one observe
that $k_+$ in the argument of the $\delta-$function can be ignored due
to the rotation invariance  and one obtains
\be{G1}
G_1=-i\varepsilon\lbrack abc]~a_{\chi}~\delta(x )
 \int dk~\frac{ k^2}
{{m_\pi^2}\left[  k^{2}-m_\pi^{2}\right]^{2}},
\ee
where we used the normalization condition \re{norm}.
The integral over the momentum $k$ can be easily computed and the result
defines the constant $D^{(1)}_0$ which reproduces
the contribution with the $\delta-$function from \re{q1loop}.

The next important observation is very simple. One can expect
that at the higher orders the non-trivial contributions to the
coefficients $D_i$ can be generated in similar way if the four-pion
subgraphs generates  the scalar products $(k\cdot p)^i$
with the highest possible power $i$ in the numerator.
Then the $i+1$-loop graph for the forward
matrix element includes the following  master integral
\be{Gi1}
G_{i+1}\sim~\frac{1}{\varepsilon^{i}}a^{i+1}_{\chi}~
q^{(0)}(\beta)* \int dk~
\frac{ k_+\delta(x p_+ -\beta k_+)}
{ \left[  k^{2}-m_\pi^{2}\right]^{2}}
 \frac{(k\cdot p)^{(i+1)}}{m_\pi^{2(i+1)}} + ...   ,
\ee
This integral can be computed in the same way as before,
 but now one has
to perform the expansion of the $\delta-$function with respect to $k_+$ in
 in order to obtain factor $k_+^i$ which is necessary in order to contract momentum
 $k$ in the scalar products. Hence the master integral can provide contribution
 with $\sim \delta^{(i)}(x) a^{i+1}_{\chi}$. At the same time, the
 integral is $UV-$divergent  and therefore the result contains the contribution
 with the maximal power of the logarithm $\sim a^{i+1}_{\chi}\ln^{i+1}[1/a_\chi]$.
 This is exactly the contribution which we want to compute.
  All this  is an illustration of the
 main idea how the higher derivatives of the $\delta-$function may appear at the
 higher orders. In order to perform the explicit calculation of the model
 independent logarithmic contributions one has to take into account
 the presence of the
 subdivergencies in the multiloop graphs. The idea of the method is quite
 general and
based on the locality of the $UV-$counterterms. For the two-loop case it
 was discussed  in Ref.~\ci{col} but can be easily extended to arbitrary
  order
 of $\chi$PT. For convenience, we provide below the technical analysis  for
 the two-loop calculation of chiral logarithms in our case.

%Let us first sketch the method of the calculation of leading two loop chiral
%logarithms. We follow closely the method described in ref.~\cite{colangelo}\
%In order to compute model independent logarithmic
%contribution ($\sim [m_\pi^2 \ln(m_\pi^2)]^2$) we have to
%calculate $1/\varepsilon^{2}$  divergencies of the two-loop diagrams.
%The corresponding divergencies can be obtained from the calculation of
%one loop diagrams. The observation is following.

The renormalized matrix element of an operator
to two-loop accuracy can be written as:
\be{ORme}
\left\langle O_{R}\right\rangle =~t^{2\varepsilon}\langle O^{(2)}%
\rangle +\frac{1}{\varepsilon}~ t^{\varepsilon
}~\langle \tilde{O}^{(1)}\rangle +\frac{1}{\varepsilon^{2}%
}\langle \tilde{O}^{(0)}\rangle
\ee
Let us clarify the rhs of eq.~\re{ORme}.
The first term $\langle O^{(2)}\rangle $ is the bare matrix element (two-loop
graph).
\begin{figure}[ptb]
\begin{center}
\includegraphics[
height=0.9105in, width=5.9453in
]%
{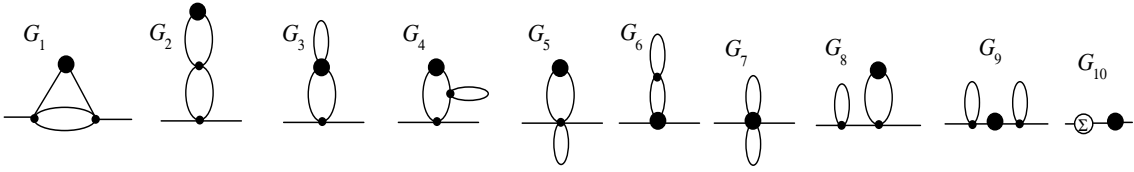}%
\end{center}
\caption{Two-loop diagrams with the two-pion operator vertex
contributing to the
%$(a\ \ln(1/a))^2$ order of the
chiral expansion of the pion parton distributions.}
\label{2loopdms}
\end{figure}
\begin{figure}
[ptb]
\begin{center}
\includegraphics[
height=0.6cm
]%
{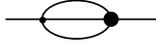}%
\end{center}
\caption{Two-loop diagram with the true four-pion operator
vertex (without tadpoles) contributing to the
%$(a\ \ln(1/a))^2$ order of the
chiral expansion of the pion parton distributions. The singular $\delta-$functions
are absent in the contribution from this graph, see explanation in the text.
}
\label{4piondiag}
\end{figure}
The second term
$\frac{1}{\varepsilon}~\langle \tilde{O}^{(1)}\rangle$
denotes subtractions of the one-loop subdivergencies
associated with the operator
vertex or pion Green functions. The one-loop reduced matrix element
$\langle \tilde{O}^{(1)}\rangle$ includes effective vertex from the
corresponding one-loop counterterm.
And the third term is proportional to the tree level matrix element
$\langle \tilde{O}^{(0)}\rangle$ of some effective operator. This term
corresponds to the local counterterm, which removes the
total divergency. The dimensionless
combination $t=\frac{\mu_\chi^{2}}{m_\pi^{2}}$ is introduced for convenience.
To two-loop accuracy the structure of divergencies of each contribution in
eq.\re{ORme} has the following form:%
\be{ORme:terms}
\langle O^{(2)}\rangle =\frac{1}{\varepsilon^{2}}g_{2}+\frac
{1}{\varepsilon}g_{1}+g_{0},~\ \ \ \langle \tilde{O}^{(1)}\rangle
=\frac{1}{\varepsilon}h_{1}+...~,
\ee
where functions $g_i$ and $h_i$ are polynomials of external momenta and
the pion mass. Inserting these expansions into \re{ORme}
 one obtains%
\be{ORme1}
\left\langle O\right\rangle _{R}&=&\frac{1}{\varepsilon^{2}}g_{2}\left[
1+2\varepsilon~\ln t+2\varepsilon^{2}\ln^{2} t%
\right]  +\frac{1}{\varepsilon^{2}}~h_{1}\left[
1+\varepsilon~\ln t+\frac{1}{2}\varepsilon^{2}\ln^{2}%
t \right]  +~...~
\\
%\ee%
%\be{ORme2}
& =&\frac{1}{\varepsilon^{2}}\left[  g_{2}+~h_{1}\right]  +\frac
{1}{\varepsilon}\left[  2g_{2}+h_{1}\right]  \ln t%
+\underset{local~terms}{\underbrace{\frac{1}{\varepsilon}\left[
...\right]  }}
%\\&
+\ln^{2} t\left[  2g_{2}+\frac{1}{2}%
~h_{1}\right]  +...~.
\ee
Now, the non-local $1/\varepsilon$ divergency proportional to $\ln t$
must cancel (locality of counterterms), that provides  the relation:
\be{g2toh1}
2g_{2}+h_{1} =0,\, \Rightarrow
g_{2} =-\frac{1}{2}~h_{1}.
\ee
i.e., the residue of the $1/\varepsilon^{2}$ is defined in terms of
the one-loop diagrams with insertions of one-loop counterterms.
 This defines completely
the coefficient in front of double logarithmic contributions:%
\be{LLcoef}
2g_{2}+\frac{1}{2}h_{1}=-\frac{1}{2}h_{1}%
\ee
Therefore, in order to define the contribution of the type
$\sim  \ln^2(m_\pi^2)$ we can  compute only $1/\varepsilon^2$ pole in
the  diagrams with insertion of the one-loop counterterms. In addition,
we have to select such structures, which can produce $\delta-$functions.
From the observation considered above, one can expect that the two-loop diagrams
can contain master integrals eq.\re{Gi1} with the power  $i=1$,
hence such diagrams are relevant for the computation
 of the coefficient $D_1$ in front of  $\delta^\prime(x)$
 from \re{QchptD} ( $I=0$ ).

At the  NNLO of $\chi$PT we have set of the two-loop diagrams
presented in Fig.~\ref{2loopdms} and Fig.~\ref{4piondiag}.
\begin{figure}[ptb]
\begin{center}
\includegraphics[
height=2cm
]%
{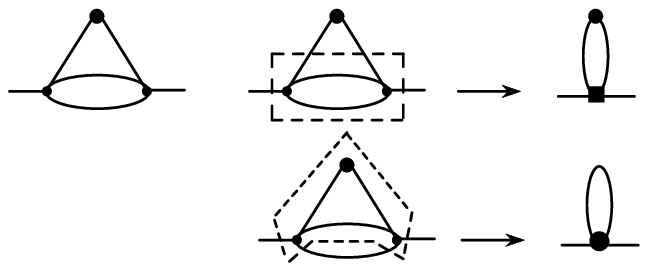}%
\end{center}
\caption{One-loop sub-divergencies of the two-loop diagram $G_1$
and corresponding counterterm. The filled circle is an operator
counter, filled square is the 4-pion Green function counterterm. }
\label{subd}
\end{figure}
Diagrams $G_3-G_9$ are
insertions of the tadpole loops to the one-loop diagrams.
From simple analysis one can conclude that these diagrams
do not contain the master integral like \re{Gi1} with the maximal power $i=2$
for scalar product $(k\cdot p)$ because of dimensional reasons:
the tadpole integral is proportional
to the pion mass $\sim m_\pi^2$ and therefore reduces the possible power $i$ by one unit.
It means that these diagrams have the same
singular structures \ as the one-loop graphs, i.e., they can have only
$\delta(x)$ terms without derivatives.  The more singular
terms may appear only from the first two diagrams $G_{1,2}$.

Consider these graphs. There are two one-loop divergent subgraphs
in these diagrams: operator vertex subgraph  and pure pion
subgraph. They are depicted in Fig.~\ref{subd}. There are two
corresponding counterterms:
\be{1lcoterms} \frac{1}{\varepsilon}~
t^{\varepsilon}~\langle \tilde {O}^{(1)}\rangle
=\frac{1}{\varepsilon}~ t^{\varepsilon}~\langle
\tilde{O}_{V}^{(1)}\rangle +\frac {1}{\varepsilon}~
t^{\varepsilon}~\langle \tilde{O}_{\pi}^{(1)}\rangle , \ee
Computation of  these contributions provides following
result%
\be{D1}
D_1=\left(  -\frac{5}{3}\right)  M_{2}^{Q}~
\ln^{2}\left[  \frac{\mu_\chi^{2}}{m_\pi^{2}}\right]  ~,
\ee
No new  singular structures of the isovector PDF $q(x)$
appear, as the isovector PDF is symmetric.

Let us briefly discuss the diagram on Fig.\ref{4piondiag}. This diagram
includes the four-pion operator vertex\footnote{We do not consider here the
operator vertices with the tadpole loops as the true multiparticle vertices}.
Such vertex can be generated either from
the operators \re{matchingL} and \re{matchingR} or from the
multiparticle  terms
which we do not write explicitly in Eqs.~(\ref{matchingL},\ref{matchingR}).
These additional multiparticle terms of the chiral operator
 include new chiral constants and do
not generate two pion vertices. Their expansion begins from the four-pion
contributions as a minimum.
We checked that the diagram Fig.~\ref{4piondiag} does not produce the contributions
with the $\delta-$functions in the case when the vertex is generated from the
operators \re{matchingL} and \re{matchingR}. This can be seen from the fact
that one pion from the operator vertex describes the external state and
therefore provides the non-trivial argument for the distribution of the
fraction $x$ that excludes occurrence $\delta(x)$. The contributions of the
multiparticle terms with the new chiral constants have been ignored for the same reason.

Now let us briefly discuss the general mechanism of generation $\delta-$singular terms.
It turned out, that such
terms appear only from the diagrams with four-pion counterterms and only in the
diagram $G_1$. We checked that the operator vertex counterterm,
provides only simple $\delta(x)$ without derivatives. The term with the
pion counterterm exactly reproduces the master integral \re{Gi1} with $i=2$.
  It also naturally explains the absence of $\delta^{\prime}(x)$ at one loop:
  the four-pion vertex is quadratically divergent and in the
corresponding  master integral can occur only simple $\delta-$function
structure considered in eq.\re{onel}.
Note also that the $\delta-$producing master integrals have always  maximal
value of the  $UV-$divergency index, i.e., power divergent terms play
the key role in the generation of the $\delta-$singular contributions.

Previous analysis can be easily extended to the three-loop diagrams.
% In this case we  have two  graphs with the two pion operator vertex
%and therefore our chiral Lagrangian for the light-cone operator remains valid.
Following the method discussed above one can again reduce
the calculation of the $a^3_\chi \ln^3 [1/a_\chi]$ contribution to the reduced
one loop diagram with effective vertex corresponding to the
insertion of the two-loop counterterms.

\begin{figure}
[ptb]
\begin{center}
\includegraphics[
height=1.5cm
]%
{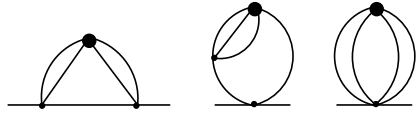}%
\caption{Three-loop diagrams with the four-pion operator vertices }%
\label{4pi3l}%
\end{center}
\end{figure}

We again have the subset of the graphs with the four-pion operator vertices
shown in Fig.\ref{4pi3l}.
In the case of isospin $I=1$ such vertices appear only from the multiparticle
part of the effective twist-2 operator with the new chiral constants.
Therefore such contribution without loss
of generality can be considered separately. In this paper we shall ignore them
what, however, cannot change our main conclusions.

\begin{figure}
[ptb]
\begin{center}
\includegraphics[
height=0.9889in,
width=5.1677in
]%
{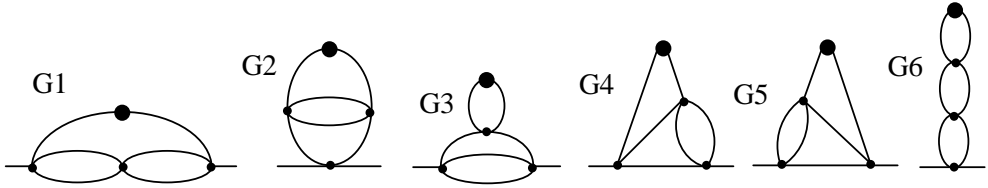}%
\caption{Three-loop diagrams which may produce $\delta-$function contributions. }%
\label{fig:3loops}%
\end{center}
\end{figure}

\begin{figure}
[ptbptb]
\begin{center}
\includegraphics[
height=0.9151in,
width=2.9288in
]%
{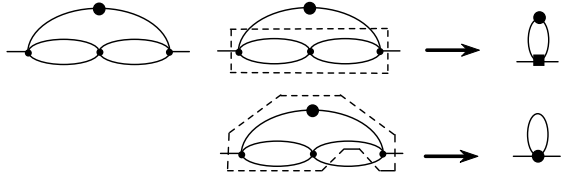}%
\caption{Diagram $G_1$ (left) and two reduced graphs with
effective vertices (right) which have to be computed in order to define
$\delta-$function logarithmic contribution. By the dashed line
we show  the two-loop subgraphs which has to
be substituted by its $UV-$counterterm.}
\label{fig:D1}%
\end{center}
\end{figure}

The set of the ``two-pion'' relevant graphs is given in Fig.~\ref{fig:3loops}.
In Fig.~\ref{fig:D1} we show the two-loop subgraphs for the case of
diagram $G_1$ and resulting one-loop graphs with effective vertices.
We do not show all
diagrams with  tadpole loops and wave function renormalization.
It is clear that such contributions can produce only $\delta-$functions
which occur in the one- and two-loop graphs.
But now we are interested in the new, more
singular  terms with the
second derivative $D_2 \delta^{(2)}(x)$ in Eq.~\re{qchptD}.
The main steps of the calculation are the same as for the two-loop case.
Again the singular contributions occur only from the reduced diagrams with
effective vertex generated by the  four-pion subgraphs
( the upper line in Fig.~\ref{fig:D1}).
The substitution of the effective counterterm vertex reduce the calculation
to the master integral \re{Gi1}.
Only the diagrams $G_1$, $G_4$ and $G_5$ provides non-trivial contribution
 to the coefficient $D_2$. The three-loops  result is %
\be{D2res}
D_2=a_\chi^{3}\ln^{3}\left[{1}/{a_\chi}\right]
~~\left\langle x^{2}\right\rangle ~\left(  -~\frac
{25}{108}\right).
\ee
where $\left\langle x^{2}\right\rangle$ denotes the moments of the PDFs in the
chiral limit
\be{mom}
\left\langle x^{n-1}\right\rangle=\int_0^1 dx\, x^{n-1}q^{(0)}(x)
\ee
%EndExpansion
%
%EndExpansion
%
%EndExpansion

It is obvious, that given analysis can be continued further order by
 order in  $\chi$PT. We do not find
any special reasons that could suppress contributions with the higher
derivatives $\delta^{(n)}(x)$ in \re{QchptD} and \re{qchptD}.
Hence, we may conclude that
structure of the chiral expansions for pion PDFs  proposed in
formulas \re{QchptD} and \re{qchptD} is established.

\section{Calculation of the singular terms in the pion GPDs }

In this section we  discuss shortly the results for the singular
contributions to the pion GPDs, counterparts of the
$\delta-$contributions to the pion PDFs, see eq.~\re{Qexp}. We shall
consider, for simplicity, the limit
\be{eq31}
t\rightarrow0,~\xi\neq0.~
\ee
Logarithmic one-loop calculation gives \cite{kiv02,man}:
\be{eq32}
H^{I=1}(x,\xi)|_{\xi\neq0,~t=0}=\mathring{H}^{I=1}(x,\xi)
\left(  ~1+a_\chi\ln\left[{1}/{a_\chi}\right]\right)
 -  a_\chi\ln\left[1/a_\chi\right]
\frac{\theta\left(  \left\vert x\right\vert
\leq\xi\right)  }{\xi}\mathring{\varphi}_{\pi}\left(  \frac{x}{\xi}\right)  .
\ee
Here the second term in the forward limit ($\xi\to 0$) tends to the
$\delta(x)$ contribution for the pion PDF
\be{eq33}
\frac{\theta\left(  \left\vert x\right\vert \leq\xi\right)  }{\xi}%
\mathring{\varphi}_{\pi}\left(  \frac{x}{\xi}\right)  \overset{\xi
\rightarrow0}{\rightarrow}\delta(x)
\ee
In order to obtain above result we used
 that pion distribution amplitude $\mathring{\varphi}_{\pi}(z)$
in the chiral limit, due to the soft-pion theorem \cite{Polyakov:1998ze},
can be represented in terms of double distribution as
\be{eq34}
\mathring{\varphi}_{\pi}(z)= {F}^{I=1}(\beta , \alpha)\, *
\delta(\alpha+\beta-z)\, ,
\ee

The isoscalar pion GPD does not have logarithmic contribution
at the one-loop order.
Calculations of the singular higher loop contribution goes along
the line described in previous section, the final result is as follows:
\begin{eqnarray}
\label{GPDsing}
{H}^{I=0}(x,\xi)&=&{H}_{reg}^{I=0}(x,\xi)
+\frac{5}{3}~
a_\chi ^{2}\ln^{2}\left[{1}/{a_\chi}\right]~\frac
{\theta\left(  \left\vert x\right\vert
\leq\xi\right)}{\xi^{2}}\ D\left(\frac{x}{\xi}\right)\, ,\\
H^{I=1}(x,\xi) & =&
H_{reg}^{I=1}(x,\xi)  -
\theta\left(  \left\vert x\right\vert
\leq\xi\right) \left\{
a_\chi\ln\left[{1}/{a_\chi}\right]
\frac{1}{\xi}\mathring{\varphi}_{\pi}\left(  \frac{x}{\xi}\right)
 -\frac{25}{108} a_\chi^3 \ln^{3}\left[{1}/{a_\chi}\right]\frac{1}{~3 \xi^{3}}
~\psi \left(  \frac{x}{\xi}\right)\right\}\, .
\end{eqnarray}
Here $D(z)$ is the so-called $D$-term \cite{Pol99}, which in the chiral limit, due to the soft-pion theorem \cite{Polyakov:1998ze},
can be expressed as:
\be{eq35}
D(z)= 2 {F}^{I=0}(\beta , \alpha)\, * \delta(\alpha+\beta-z)\, ,
\ee
and function $\psi(z)$ is expressed in terms of double distribution as follows:
\be{eq36}
\psi\left(  z\right)  = \lim_{\eta\rightarrow 1}\frac{\partial^2}{\partial \eta^2}\eta {F}^{I=1}(\beta,\alpha)\, * \beta\ \delta\left(
\alpha+\beta\eta -z\right)\,.
\ee
One can easily check that in the forward limit ($\xi\to 0$)
 the equation (\ref{GPDsing}) is reduced to the
result for the pion PDFs (\ref{Qexp}).

The singular terms of GPDs (\ref{GPDsing}) result in the contributions to the amplitudes for hard exclusive
processes of the following type:

\be{eq37}
\sim \left(\frac{a_\chi \ln(1/a_\chi)}{\xi}\right)^n, \ \ \ n=1,2,3\ldots
\ee
This clearly demonstrates that any finite order of chiral corrections to
physical observables at $\xi \sim a_\chi \ln(1/a_\chi)$
gives contribution which is not suppressed by small chiral parameter.
 Therefore in this region of kinematical variable
$\xi$ one needs to perform infinite order resummation of the
chiral perturbation theory.

\section{Discussion}

In present paper we have demonstrated that chiral expansion of the pion PDFs
includes the contributions of the $\delta-$function and its derivatives.
We have also computed the leading non-analytic contributions to the
coefficients $D_{1,2,3}$ in \re{QchptD} and \re{qchptD}.
The results read:
\begin{align}
Q(x)  & =Q^{reg}(x)+
~\left(  -\frac{5}{3}\right)  M_{2}^{Q}~a_\chi^{2}
\ln^{2}\left[  {1}/{a_\chi}\right]~\delta^{\prime
}(x) +\mathcal{O}(a_\chi^{3}),\label{Qexp}
\\
q(x)  & =q^{reg}(x) -a_\chi\ln\left[
{1}/{a_\chi}\right]\ \delta(x) +\left\langle
x^{2}\right\rangle \left(
-\frac{25}{108}\right)~a_\chi^{3}
\ln^{3}\left[ {1}/{a_\chi}\right]  ~\delta^{\prime\prime
}(x)  +\mathcal{O}%
(a_\chi^{4}),
\end{align}

It is also clear that higher order terms of the chiral
Lagrangian for pion dynamics $\mathcal{L}_{6,8...}$ and  for
the light-cone operators
can also produce $\delta-$singular terms but they are suppressed by powers of the
small chiral parameter with respect to {\it leading chiral logarithms}
($\sim [a_\chi\ln(1/a_\chi)]$) by additional powers of $a_\chi$.
%%%%%%%%%%%%%%%%%%%%%%%%%%%%%%%%%%%%%%%%%%%%%%%%%%%%%%%%%%%%%%%%%%%%%%

 From the obtained results we can conclude that in the region of small
momentum fraction $x$
%is not correctly described by    !changed
application of
the standard
chiral expansion faces with the difficulties.
%In this region     !changed
In this region one cannot truncate the perturbation expansion and
hence to perform
some sort of resummation. This
reorganization of the chiral expansion may  combine the singular
contributions to some generating function $f(z)$:
\begin{equation}
\sum D_{n}~\varepsilon^{n}\delta^{(n)}(x)=f(x~/\varepsilon)\label{theta}%
\end{equation}
where small parameter $\varepsilon=a_\chi\ln\left[  {1}/{a_\chi}\right]  $.
The function $f(z)$\ is some stable
in the chiral limit
function, which, probably, can be defined in the whole real axes
$-\infty\,<z<\infty$ and generates contributions to the moments in
the form
\be{fmom}
\int_{-1}^{1}dx~x^{n-1}~f(x~/\varepsilon)\approx \varepsilon^{n}%
\int_{-\infty}^{\infty}dz~z^{n-1}~f(z)\simeq
\varepsilon^{n}~f_{n}=\varepsilon^n\ (-1)^n (n-1)!\ D_{n-1}.
\ee
The obvious conclusion which follows from this discussion is that any finite
order calculation of the PDF moments can not provide enough information in
order to reconstruct PDF in the whole region of variable $x$. One has to
perform resummation of singular $\delta-$function terms but this is equivalent
to calculation of all orders $\chi$PT.

Of course, such resummation is very difficult task. Even if we restrict our
consideration only by the leading non-analytic terms we must compute
straightforwardly order by order all terms $D_i$ which is not possible.
Hence one has to introduce some simplifications or models
in order to achieve the goal.
As an example, let us provide here the result of such simplified
resummation in the leading
order of the $1/N$ expansion, where
$N$ is number of pions ($N=3$ for the real two flavor QCD).
This expansion is obtained by the generalization of the  $SU(2)\times SU(2)=O(4)$
chiral model to the $O(N+1)$ model.
The latter model can be solved in the large $N$ limit. The result of
the calculations for the isovector PDF is
\footnote{Details of the calculations and discussion
will be given elsewhere \cite{inprep}}:
\begin{equation}
q(x)\simeq q_{reg}(x)-\frac{2}{N}\ \theta(N \varepsilon-|x|)\int_{|x|/(N \varepsilon)}^{1}\frac{d z}{z}~~q^{(0)}(z
)\, .
\end{equation}
[Note, that parametrically
$\varepsilon=1/2 a_\chi\ln\left[{1}/{a_\chi}\right]\sim 1/N$].
 This example of chiral resummation shows clearly that calculation
of any finite order of $\chi$PT contributions to the Mellin moments of PDF does
not allow to restore correctly the distribution function for
$x<3/2a_\chi\ln\left[{1}/{a_\chi}\right]$ (numerically for the physical pion mass $x<0.09$
that is not very small).
%%%%%%%%%%%%%%%%%%%   !changes
 For the chiral extrapolation of the lattice data for PDFs  this implies that
 usage of the moments without resummation
%such extrapolation
 (see e.g. \cite{lattice}) can provide us the incomplete information
about parton distributions in the region of $x\sim a_\chi \ln[1/a_\chi]$.
In other words, one can
use $\chi$PT in order to compute systematically the local moments of the PDFs but one cannot use these
moments in order to describe the PDFs in the region of the small momentum fraction where all orders of the
$\chi$PT contribute to the same accuracy.

%%%%%%%%%%%%%%%%%%%   !changes
%After all it is not surprising as in the lattice calculations
%one is able to access only observables which can be defined in
%Euclidean space, whereas the parton distributions "live"
%essentially  in Minkowski space.

\section{Acknowledgments}
The
work is supported
by the Sofja Kovalevskaja Programme of the Alexander von Humboldt
Foundation, the Federal Ministry of Education and Research and the
Programme for Investment in the Future of German Government.

\section{Appendix. Twist-2 operators and the matrix elements}
In this section we briefly describe the  definitions and some technical
details used in the paper. We introduce two light-like vectors $n,\, \bar n$:
\be{llv}
n^2=\bar n^2=0,  n\cdot\bar n=1,\,\, a_{+}=a\cdot n.
\ee
There exist two QCD quark light-cone operators of twist-2:%
\be{PLR}
P_{R}=\frac{1}{2}\left(  1-\gamma_{5}\right)  ,~~P_{L}=\frac{1}{2}\left(
1+\gamma_{5}\right)  ,
\ee
\be{OLR}
\left[  O_{R}\right]  _{fg}  & =\bar{q}_{g}\left(  \frac{1}{2}\lambda
~n\right)  ~\gamma_{+}P_{R}~q_{f}\left(  -\frac{1}{2}\lambda~n\right)  ,\\
\left[  O_{L}\right]  _{fg}  & =\bar{q}_{g}\left(  \frac{1}{2}\lambda
~n\right)  ~\gamma_{+}P_{L}~q_{f}\left(  -\frac{1}{2}\lambda~n\right)  ,
\ee
where indexes $f,g$ stand for flavor. These operators transform
under global chiral rotations  as
\be{Chrot}
  O_{L}\rightarrow V_L   O_{L} V_L^\dag
\,,\,\,
 O_{R}\rightarrow V_R   O_{R} V_R^\dag\, .
\ee
In $\chi$PT these QCD operators are described by effective chiral operator with
unknown chiral constants. In the pure pion sector ($U$ as usually denotes pion field)
one finds \ci{kiv02,man,che}
\be{matchingL}
O^L_{fg}(\lambda)&=&
-\frac{i F_\pi^2}{4}\
\mathcal{F}(\beta , \alpha)*
 \left[U\left(\frac{\alpha+\beta}{2}\lambda n\right)n\cdot\partialboth
U^\dagger\left(\frac{\alpha-\beta}{2}\lambda n\right)\right]_{fg} \, ,
\\
O^R_{fg}(\lambda)&=&
-\frac{i F_\pi^2}{4}\  \mathcal{F}(\beta , \alpha)*
 \left[U^\dagger\left(\frac{\alpha+\beta}{2}\lambda n\right)n\cdot\partialboth
U\left(\frac{\alpha-\beta}{2}\lambda n\right)\right]_{fg}\, .
\label{matchingR}
\ee
where by asterisk we denote the integral convolution with
respect to $\beta$ and $\alpha$:
\be{asterisk}
\mathcal{F}(\beta , \alpha)*O(\beta,\alpha)
\equiv
 \int_{-1}^1 d\beta \int_{-1+|\beta|}^{1-|\beta|} d\alpha\
\mathcal{F}(\beta , \alpha)\, O(\beta,\alpha)\, .
\ee
Here $\mathcal{F}(\beta , \alpha)$ represents the real generating function
of the tower of
low-energy constants and $\partialboth_\mu$ denotes combination of derivatives
$\stackrel{\rightarrow}{\partial_\mu}-\stackrel{\leftarrow}{\partial_\mu}$.
%%%%%%%%%%                !changed
The non-local structure of the operators \re{matchingL} and \re{matchingR} implies
that the light-cone distance $\lambda$ has chiral counting $\lambda\sim {\cal O}(p^{-1})$
as suggested in \ci{kiv02}.
Then the combination $\lambda \partial_+$ is dimensionless that provides the non-locality
with respect to light-cone direction.
%%%%%%%%%%%%%%%%%%%%%%%%%%%%%%%%%%%%%%%%%%
Important to note that given Lagrangian is not complete: the expressions
 \re{matchingL} and \re{matchingR} describe correctly only operators
 vertices with two attached pions (including the tadpoles loops).
The low-energy constants $\mathcal{F}(\beta , \alpha)$ are characteristics
of the structure of the pion, they are not determined in the effective
field theory. According to the isospin $I=0,1$ one can construct two
independent functions:%
\be{def:FI}
F^{I=0}\left[  \beta,\alpha\right]   & =\frac12(\mathcal{F}\left[  -\beta
,\alpha\right]  -\mathcal{F}\left[  \beta,\alpha\right])  ,\\
F^{I=1}\left[  \beta,\alpha\right]   & =\frac12(\mathcal{F}\left[  -\beta
,\alpha\right]  +\mathcal{F}\left[  \beta,\alpha\right] ) ,
\ee
which are convenient to describe the pion matrix elements. Pion PDFs are
defined as
\be{def:PDFs}
\int\frac{d\lambda}{2\pi}e^{-ip_+\,x\lambda}\left\langle \pi^{b}(p^{
})\left\vert \text{tr}\left[  \tau^{c}O_{L+R}(\lambda)\right]  \right\vert
\pi^{a}(p)\right\rangle  & =4i\varepsilon\lbrack abc]q(x),\\
\int\frac{d\lambda}{2\pi}e^{-ip_+\, x\lambda}\left\langle \pi^{b}(p)
\left\vert \text{tr}\left[  O_{L+R}(\lambda)\right]  \right\vert \pi
^{a}(p)\right\rangle  & =2\delta^{ab}Q(x)
\ee
that in the chiral limit can be written as
\be{}
2 \int_{-1+\left\vert \beta\right\vert }^{1-\left\vert \beta\right\vert }%
d\alpha~F^{I=0}(\beta,\alpha)~  & =\left[  \theta(\beta
)~q(\beta)-\theta(-\beta)~\bar{q}(-\beta)\right]  =Q(\beta),\\
\int_{-1+\left\vert \beta\right\vert }^{1-\left\vert \beta\right\vert }%
d\alpha~F^{I=1}(\beta,\alpha)~  & =\theta(\beta)~q(\beta)+\theta(-\beta
)~\bar{q}(-\beta)=q(\beta).
\ee{SumR}
The more general functions GPDs are defined as
\be{def:GPDs}
\int\frac{d\lambda}{2\pi}e^{-iP_+\, x\lambda}\left\langle \pi^{b}(p^{\prime
})\left\vert \text{tr}\left[  \tau^{c}O_{L+R}(\lambda)\right]  \right\vert
\pi^{a}(p)\right\rangle  & =4i\varepsilon\lbrack abc]H^{I=1}(x,\xi,t),\\
\int\frac{d\lambda}{2\pi}e^{-iP_+\,x\lambda}\left\langle \pi^{b}(p^{\prime
})\left\vert \text{tr}\left[  O_{L+R}(\lambda)\right]  \right\vert \pi
^{a}(p)\right\rangle  & =2\delta^{ab}H^{I=0}(x,\xi,t)
\ee
with~$P=\frac{1}{2}(p+p^{\prime}),~\xi=-\frac{(p^{\prime}-p)_+}
{(p^{\prime}+p)_+},~t=(p^{\prime}-p)^{2}.~$
In the forward limit $\xi\rightarrow0,\, t\rightarrow 0 $:%
\be{fwd}
H^{I=1}(x,0,0)=q(x),~\ H^{I=0}(x,0,0)=Q(x).
\ee
And the pion distribution amplitude is given
\be{phiDA}
\delta^{ab}F_\pi~\phi_{\pi}(u)  & =\frac{i}{4}\int\frac{d\lambda}{2\pi}e^{-iu(p.n)\lambda
/2}\left\langle \pi^{a}(p)\left\vert \text{tr}\left[  \tau^{b}O_{R-L}%
(\lambda)\right]  \right\vert 0\right\rangle ,
\ee
Defined  distribution amplitude satisfies normalization
\be{normDA}
\int_{-1}^{1}du~~\phi_{\pi}(u)=1.
\ee

\end{document}